\documentclass[aps,twocolumn,groupedaddress,showpacs]{revtex4}
\usepackage[dvips]{graphicx}
\usepackage[]{caption}
\usepackage{amsmath}
\usepackage{amssymb}
\begin{document}

\bibliographystyle{prsty}

\title{Simple Equations for Cosmological Matter and Inflaton Field Interactions}
\author{E. Rebhan and T.~J. Battefeld$^{1}$}
\affiliation{Institut f{\"u}r Theoretische Physik,
  Heinrich--Heine--Universit{\"a}t, D--40225 D{\"u}sseldorf, Germany,\\
  \\
  $^1$Present address: Physics Department, Brown University,
  Providence RI 02912 USA.}

\date{\today}

\begin{abstract}
  A simple model for the reheating of the universe after inflation is
  studied in which an essentially inhomogeneous scalar field
  representing matter is coupled to an essentially homogeneous scalar
  inflaton field. Through this coupling, the potential determining the
  evolution of the inflaton field is made time-dependent. Due to this
  the frequency of parametric resonance becomes time dependent, making
  the reheating process especially effective.

  All fields including the gravitational field are initially
  simplified by expanding each in terms of the respective homogeneity
  or inhomogeneity.  Employing only the lowest order of this
  expansion, we space-average, and introduce all independent averages
  as new variables. This leads to a hierarchy of equations for the
  spatial moments of the fields and their derivatives. A small
  expansion parameter permits a truncation to lowest order, yielding a
  closed system of 5 coupled nonlinear first order differential
  equations.

  For a parabolic potential, the energy densities of the matter and
  inflaton fields oscillate chaotically around each other from the end
  of inflation until they reach extremely small values. The average
  period of preponderance of one of the two continuously increases in
  this process. We discuss that this may provide one clue to a
  solution of the coincidence problem.

  For a Mexican hat potential we can easily obtain and understand
  dynamical symmetry breaking.

\end{abstract}

\maketitle

\section{Introduction}

Inflation has become an almost indispensable ingredient of cosmology
because so far it is the only means by which several problems of
cosmology, like the horizon, the flatness or the structure formation
problem can be solved. Inflation can be achieved by an inflaton field
\cite{linde,lyth,brand3}, i.e. a homogeneous field which exerts
negative pressure and, as the ground state of a quantum field, has no
particles. In its simplest version it is a scalar field which, for
simplicity, will also be used here.

On one hand there must be an end to inflation, and on the other, at
the end or after inflation there must be formation and heating and/or
reheating of matter which, if initially present, has been extremely
diluted and cooled down by inflation. In order for inflation to come
to an end, the inflaton field must loose much of its energy while for
its formation and (re-)heating, matter must acquire energy.  In
principle this can happen independently: Matter can extract energy
from the gravitational field, and energy of the inflaton field can go
to the gravitational field which effectively amounts to a transfer of
energy from the inflaton to the matter field.  These processes can be
independent even if they occur at the same time.  Inflation scenarios
employing only an independent inflaton field are of this kind.

The most convincing inflation scenarios appear to be those in which
energy is transferred from the inflaton to the matter field through
direct coupling. In this case it was shown that reheating is
especially effective due to the occurrence of parametric resonance
\cite{linde2,linde3,linde4,brand,brand2}, if a field description for
matter is used. The scenario employed in this paper is of this type.
We observe chaotic behavior which is expected, since there is a close
relationship between chaos and parametric resonance \cite{joras}.  A
different approach studying this relationship using lattice
simulations was considered in \cite{felder}.

\section{Model}

It will be assumed that the dynamics of the inflaton field $\varphi$
is determined by a potential of the form
\begin{equation}
 V(\varphi){=}V_{0}+\frac{m^{2}\varphi^{2}}{2}
 + \lambda\frac{\varphi^{4}}{4}
\end{equation}
which is re-normalizable in a quantum field theory. We incorporate
matter by introducing a scalar field $\chi$, though this can only be a
crude reflection of reality.  An interaction Lagrangian of the form
\begin{equation}
 {\mathcal L}_{\text{I}}{=}\alpha
\frac{\chi^{2}\varphi^{2}}{2}
\end{equation}
is assumed. A potential $V=V(\varphi ,\chi)$ containing the
interaction ${\mathcal L}_{\text{I}}$ is
\begin{equation}                                                \label{4}
  V(\varphi ,\chi){=} V_0{+}\alpha\left(\chi ^2{-}\chi _c^2\right)
  \frac{\varphi^2}{2}+\lambda\frac{\varphi ^4}{4}+ \frac{m_\chi ^2\chi ^2}{2}\,.
\end{equation}
In the case that all constants $V_{0},\alpha,\chi_{c},\lambda$ and
$m_{\chi}$ are different from zero, the potential $V(\varphi,\chi)$
has, in its dependence on $\varphi$, for given $\chi>\chi_{c}$ a
parabola-like shape and for $\chi<\chi_{c}$ a sombrero-like shape
(cf.~Fig.~\ref{pic:2}). The term ${\sim}\varphi^{2}$ is implicitly
time-dependent, enabling dynamical symmetry breaking as expected from
a Higgs field.  In addition, setting $V_{0}{=}0$, $\chi_{c}{=}0$ and
$m_{\chi}{=}0$, as a second case
\begin{equation}                                             \label{4*}
V(\varphi ,\chi)=\alpha\frac{\chi ^2\varphi^2}{2}+\lambda\frac{\varphi ^4}{4}
\end{equation}
will be considered.

The model thus introduced resembles hybrid inflation models in that it
employs two coupled scalar fields.  However, it differs from them
because the second field is not an inflaton field but an inhomogeneous
field for matter.

Concerning the space dependence it is assumed that the inflaton field
starts out as and remains an essentially homogeneous scalar field
while the matter field starts out and remains essentially
inhomogeneous,
\begin{equation}                                             \label{1}
\varphi=\varphi _{\text{h}}(t)
+\varepsilon \tilde{\varphi}(\vec{x},t)\,,\qquad
\chi=\varepsilon\chi _{\text{h}}(t)+\tilde{\chi}(\vec{x},t)
\end{equation}
\noindent with $\varepsilon\ll 1$ and
$\overline{\tilde{\varphi}}{=}\overline{\tilde{\chi}}{=}0$, where
$\overline{f}{=}\int_{V}f\, d^{3}x/V$ and $V$ is a volume large enough
to contain many particles of the field $\tilde{\chi}$. Note that due
to the interaction of the fields, $\varphi$ will get an inhomogeneous
contribution even if it starts out almost completely homogeneous,
while the opposite holds for $\chi$.

The purpose of this paper is no study of fluctuations. We will,
therefore, concentrate on average properties, inhomogeneities only
being attributed to the field representation of matter. They are
necessary for obtaining positive pressure but are intended to
represent a homogeneous matter distribution on large scales. Thus
inhomogeneities occur on a much smaller scale than the one of
fluctuations used for studying structure formation.  As a consequence,
in an expansion with respect to $\epsilon$ we shall be satisfied with
lowest order results.  Therefore, and because perturbations of the
coordinates would only result in second order effects \cite{liddle},
we can use the standard coordinates of an unperturbed space-time for
$\vec{x}$ and $t$. It should be noted that the following approach is
neither covariant, nor gauge invariant.  However, this does not matter
because of the just mentioned reasons and the focus on physically
meaningful quantities.

The lowest order results which will finally be obtained from an
expansion with respect to $\epsilon$ could easily be guessed.
Nevertheless, for the reason of clarity they will be derived
explicitly.

\section{Derivation of the basic equations }

Owing to the spatial fluctuations of the fields $\varphi$ and $\chi$,
fluctuations of the gravitational field will occur.  A time dependent
and spatially fluctuating gravitational field can be described by a
local scale factor
\begin{equation}                                             \label{0*}
 a(\vec{x},t)=a_{\text{h}}(t) +\epsilon\tilde{a}(\vec{x},t)
\end{equation}
and a local Hubble parameter \cite{liddle}
\begin{equation}                                             \label{0}
H(\vec{x},t)=\frac{\dot{a}(\vec{x},t)}{a(\vec{x},t)}
\end{equation}
(with $\dot{a}{=}\partial a/\partial t$ etc.), which
can be decomposed according to
\begin{equation}                                             \label{1*}
H(\vec{x},t)=H_{\text{h}}(t)+\epsilon \tilde{h}(\vec{x},t)
\end{equation}
with $\overline{\tilde{h}}{=}0$. Assuming minimal coupling, the
fields $\varphi$ and $\chi$ must satisfy modified
Klein-Gordon-equations
\begin{eqnarray}
&&\ddot{\varphi}+3H\dot{\varphi}-\frac{\nabla^2\varphi}{a^2}
+\frac{\partial V(\varphi,\chi)}{\partial\varphi}=0\,,\label{2}\\
&&\ddot{\chi}+3H\dot{\chi}-\frac{\nabla^2\chi}{a^2}+
\frac{\partial V(\varphi ,\chi)}{\partial\chi}=0 \,,    \label{3}
\end{eqnarray}
and $H(\vec{x},t)$ must satisfy the Raychaudhuri equation
\cite{raychaurdhuri1}--\cite{raychaurdhuri2}
\begin{eqnarray}                                             \label{3*}
\dot{H}{+}H^{2}&=&{-}\frac{4\pi}{3m^{2}_{\text{P}}}
(\rho _\varphi{+}\rho_\chi{+}
3p_\varphi{+}3p_\chi)\nonumber\\
&&-\frac{\Delta(p_{\varphi}+p_{\chi})}
{3a^{2}(\rho _\varphi{+}\rho_\chi{+}p_{\varphi}{+}p_{\chi})}\,.
\end{eqnarray}
In this the densities $\rho _\varphi$ and $\rho_\chi$ and the
pressures $\rho _\varphi$ and $\rho_\chi$ are given by
\begin{eqnarray}
 \rho _\varphi&=&\frac{\dot{\varphi}^{2}}{2}+
\frac{(\nabla\varphi)^{2}}{2a^{2}}
 +V_{\varphi}\,,\qquad
 \rho _\chi=\frac{\dot{\chi}^{2}}{2}+\frac{(\nabla\chi)^{2}}{2a^{2}}
 +V_{\chi}\,,\nonumber\\[-\jot]
 \label{a}\\[-\jot]
 p_\varphi&=&\frac{\dot{\varphi}^{2}}{2}
-\frac{(\nabla\varphi)^{2}}{6a^{2}}
 -V_{\varphi}\,,\qquad
  p_\chi=\frac{\dot{\chi}^{2}}{2}
-\frac{(\nabla\chi)^{2}}{6a^{2}}-V_{\chi}\,,
  \nonumber
\end{eqnarray}
where, with proper distribution $V=V_{\varphi}+V_{\chi}$ of the
potential (\ref{4}),
\begin{displaymath}
 V_{\varphi}=V_0{-}\alpha\chi _c^2\frac{\varphi^2}{2}
 +\lambda\frac{\varphi ^4}{4}\,,\qquad
 V_{\chi}=\alpha\chi ^2\frac{\varphi^2}{2}+\frac{m_\chi ^2\chi ^2}{2}\,.
\end{displaymath}

From Equation (\ref{0}) to lowest order in $\epsilon$ we get
\begin{equation}                                        \label{3**}
  H_{\text{h}}(t)=\frac{\dot{a}_{\text{h}}(t)}{a_{\text{h}}(t)}\,.
\end{equation}
Introducing Eqs.~(\ref{4}) and (\ref{1}) in Eqs.~(\ref{2})--(\ref{3})
and space-averaging Eq.~(\ref{2}), to lowest order in
$\epsilon$ we obtain
\begin{eqnarray}
\ddot{\varphi}_{\text{h}}+3H_{\text{h}}\dot{\varphi}_{\text{h}}
+\alpha\left(\overline{\tilde{\chi}^2}-\chi_c^2\right)
\varphi_{\text{h}}
+\lambda\varphi _{\text{h}} ^3&=&0\,,\label{5}\\
\ddot{\tilde{\chi}}+3H_{\text{h}}\dot{\tilde{\chi}}
-\frac{\nabla ^2\tilde{\chi}}{a_{\text{h}}^2(t)}+
\alpha\varphi _{\text{h}}^2\tilde{\chi}+m_\chi ^2\tilde{\chi}
&=&0\,.                   \label{6}
\end{eqnarray}

From Eqs. (\ref{a}) with (\ref{1}) we get
\begin{eqnarray*}
  \rho_\varphi&=&\rho_{\varphi 0}(t){+}
  \epsilon\tilde{\rho}_\varphi(\vec{x},t)\,,\quad
  \rho_\chi=\rho_{\chi 0}(t){+}
  \epsilon\tilde{\rho}_\chi(\vec{x},t)\,,\\
 p_\varphi&=&p_{\varphi 0}(t){+}
  \epsilon\tilde{p}_\varphi(\vec{x},t)\,,\quad
  p_\chi=p_{\chi 0}(t){+}
  \epsilon\tilde{p}_\chi(\vec{x},t)
\end{eqnarray*}
which defines $\tilde{\rho}_\varphi$, $\tilde{\rho}_\chi$,
$\tilde{p}_\varphi$, $\tilde{p}_\chi$ and
\begin{eqnarray}
  \rho_{\varphi 0}&=&\frac{\dot{\varphi}^{2}_{\text{h}}}{2}{+}
  V_{\varphi}(\varphi_{\text{h}}),\quad
  \rho_{\chi 0}=\frac{\dot{\tilde{\chi}}^{2}}{2}{+}
  \frac{(\nabla \tilde{\chi})^{2}}{2a_{\text{h}}^{2}(t)}
  {+}V_{\chi}(\varphi_{\text{h}},\tilde{\chi})\,,
  \nonumber\\[-\jot]
 \label{a*}\\[-\jot]
  p_{\varphi 0}&=&\frac{\dot{\varphi}^{2}_{\text{h}}}{2}{-}
  V_{\varphi}(\varphi_{\text{h}})\,,\quad
  p_{\chi 0}=\frac{\dot{\tilde{\chi}}^{2}}{2}{-}
  \frac{(\nabla \tilde{\chi})^{2}}{6a_{\text{h}}^{2}(t)}
  {-}V_{\chi}(\varphi_{\text{h}},\tilde{\chi})\,.\nonumber
\end{eqnarray}
With this and Eq.~(\ref{1*}), space-averaging Eq.~(\ref{3*})
yields
\begin{equation}                                       \label{3+}
\dot{H}_{\text{h}}(t){+}H_{\text{h}}^{2}(t)=
{-}\frac{4\pi}{3m^{2}_{\text{P}}}\big(\rho_{\varphi
  0}{+}\overline{\rho_{\chi 0}}{+}
3p_{\varphi 0}{+}3\overline{p_{\chi 0}}\big)
\end{equation}
to lowest order in $\epsilon$. Writing
\begin{eqnarray}
\nonumber \ddot{\varphi}_{\text{h}}+3H_{\text{h}}(t)\dot{\varphi}_{\text{h}}+
\frac{\partial\big(V_{\varphi}(\varphi_{\text{h}})+\overline{V_{\chi}
  (\varphi_{\text{h}},\tilde{\chi})}\big)}{\partial\varphi_{\text{h}}}=0
\end{eqnarray}
for Eq.~(\ref{5}) and using the equations $\rho_{\varphi
  0}{+}p_{\varphi 0}{=}\dot{\varphi}^{2}_{\text{h}}$ and
$\dot{\rho}_{\varphi
  0}{=}\dot{\varphi}_{\text{h}}\ddot{\varphi}_{\text{h}}
{+}(dV_{\varphi}(\varphi_{\text{h}})/d\varphi_{\text{h}})
\dot{\varphi}_{\text{h}}$ following from Eqs. (\ref{a*}) one obtains
\begin{equation}                              \label{dorrhoa}
 \dot{\rho}_{\varphi 0}+3H_{\text{h}}(t)(\rho_{\varphi 0}+p_{\varphi 0})
 +\frac{\partial\overline{V_{\chi}(\varphi_{\text{h}},\tilde{\chi})}}
 {\partial \varphi_{\text{h}}}\dot{\varphi}_{\text{h}}=0\,.
\end{equation}
Writing Eq. (\ref{6}) as
\begin{eqnarray}
\nonumber \ddot{\tilde{\chi}}+3H_{\text{h}}\dot{\tilde{\chi}}
-\frac{\nabla ^2\tilde{\chi}}{a_{\text{h}}^2}+\frac{\partial V_{\chi}
(\varphi_{\text{h}},\tilde{\chi})}{\partial\tilde{\chi}}=0\,,
\end{eqnarray}
multiplying
it with $\dot{\tilde{\chi}}$, space-averaging it and finally using
\begin{eqnarray}
\nonumber \overline{\dot{\tilde{\chi}}\nabla^{2}\tilde{\chi}}=
-\overline{\nabla\dot{\tilde{\chi}}\cdot\nabla\tilde{\chi}}
\end{eqnarray}
 which
follows from partial integration, we get
\begin{displaymath}
  \frac{d}{dt}\frac{\overline{\dot{\tilde{\chi}}^{2}}}{2}
 +3H_{\text{h}}(t)\overline{\dot{\tilde{\chi}}^{2}}+
  \frac{\overline{\nabla\dot{\tilde{\chi}}
\cdot\nabla\tilde{\chi}}}{a_{\text{h}}^{2}(t)}
+\overline{\frac{\partial V_{\chi}(\varphi_{\text{h}},\tilde{\chi})}
{\partial\tilde{\chi}}\dot{\tilde{\chi}}}=0\,.
\end{displaymath}
Combining this with the equations
\begin{eqnarray*}
 \dot{\overline{\rho_{\chi 0}}}&=&\frac{d}{dt}
\frac{\overline{\dot{\tilde{\chi}}^{2}}}{2}
 + \frac{\overline{\nabla\dot{\tilde{\chi}}\cdot
\nabla\tilde{\chi}}}{a_{\text{h}}^{2}(t)}
 -H_{\text{h}}(t)\frac{\overline{(\nabla\tilde{\chi})^{2}}}
 {a_{\text{h}}^{2}(t)}
+\frac{\partial \overline{V_{\chi}(\varphi_{\text{h}},\tilde{\chi})}}
{\partial\varphi_{\text{h}}}\dot{\varphi}_{\text{h}}\\
&&+\overline{\frac{\partial V_{\chi}(\varphi_{\text{h}},\tilde{\chi})}
{\partial\tilde{\chi}}\dot{\tilde{\chi}}}
\end{eqnarray*}
and $\overline{\rho_{\chi 0}}{+}\overline{p_{\chi 0}}{=}
\overline{\dot{\tilde{\chi}}^{2}}{+}
\overline{(\nabla\tilde{\chi})^{2}}/(3a_{\text{h}}^{2}(t))$ following
from (\ref{a*}) we get
\begin{equation}                                     \label{dorrhob}
 \dot{\overline{\rho_{\chi 0}}}+3H_{\text{h}}(t)
(\overline{\rho_{\chi 0}}+\overline{p_{\chi 0}})
 -\frac{\partial\overline{V_{\chi}(\varphi_{\text{h}},\tilde{\chi})}}
 {\partial \varphi_{\text{h}}}\dot{\varphi}_{\text{h}}=0\,.
\end{equation}
Addition of Eqs. (\ref{dorrhoa}) and (\ref{dorrhob}) leads to
the equation
\begin{eqnarray}
 \dot{\rho}_{0}+3H_{\text{h}}(t)(\rho_{0}+p_{0})&=&0  \label{enc}\\
\noalign{\noindent with}
\rho_{0}=\rho_{\varphi 0}+\overline{\rho_{\chi 0}}\,,\qquad
&&p_{0}=p_{\varphi 0}+\overline{p_{\chi 0}}\,.
\end{eqnarray}
It means that to lowest order in $\epsilon$ the total energy of matter
and inflaton field is conserved. It is well known that together with
Eq.  (\ref{enc}) the equation
\begin{equation}                                      \label{fried}
 H_{\text{h}}^{2}(t)+\frac{k}{a_{\text{h}}^2(t)}
=\frac{8\pi}{3m_{\text{P}}^2}\,\rho_0
\end{equation}
with $k{=}1$ or $0$ or ${-}1$ is equivalent to Eq.~(\ref{3+}) whence
we can replace the latter with Eq.~(\ref{fried}).

To lowest order in $\epsilon$ we are thus left with the set of
equations (\ref{3**})--(\ref{6}) and (\ref{fried}). They constitute a
self contained system of four equations for the four unknown
quantities $a_{\text{h}}(t)$, $H_{\text{h}}(t)$,
$\varphi_{\text{h}}(t)$ and $\tilde{\chi}(\vec{x},t)$ which completely
decouples from the first order perturbations.  In the following for
simplicity we will set $a_{\text{h}}\to a$ and $H_{\text{h}}\to H$.

Equations of a similar kind were solved in \cite{brand3,brand2} by
means of a Fourier analysis, invoking quantum field theory and
introducing several approximations. In the following a quite different
approach will be presented. This allows for a classical treatment
which gets along with minor approximations and will enable a simple
numerical treatment.

The method chosen here consist in replacing Eq.~(\ref{6}), the only
partial differential equation left in the system, by a set of ordinary
differential equations for averaged quantities. In certain aspects it
resembles the momentum method used for solving the Boltzmann equation,
the main difference being that in this paper moments are built in
ordinary space instead of momentum space. Like in Boltzmann theory,
the quality of the approximation can be improved by including moments
of ever increasing order -- a typical example is provided by the
thirteen, twenty-one or twenty-nine moment approximation in Plasma
Physics \cite{balescu}. A tremendous advantage over the momentum
method in Boltzmann theory consists in a much better quality of the
approximation provided by a truncation to low orders. This is enabled
by the fact that each additional moment entering the equations is
multiplied by the small quantity $1/a^{2}$ whence indirectly a moment
of order $n$ couples to the moments of lowest order with a factor
$1/a^{2n}$.

Defining
\begin{equation}
 x=\tilde{\chi}\,,\quad x_n=(\nabla ^nx)^2 \,,\quad y_n
 =(\nabla ^n\dot{x})^2
\end{equation}
where $\nabla^{3} x=\nabla\Delta x$ etc. one obtains
\begin{displaymath}
\ddot{\overline{x_n}}=\ddot{\overline{\left(\nabla ^nx\right)^2}}
=2\overline{(\nabla ^n\dot{x})^2}+2\overline{(\nabla ^nx)
(\nabla ^n\ddot{x})}\,.
\end{displaymath}
Inserting $\ddot{x}{=}\ddot{\tilde{\chi}}$ from Eq.~(\ref{6}) in this
yields
\begin{eqnarray*}
\nonumber \ddot{\overline{x_n}}&=&2\overline{y_n}+2\Big[-3H
\overline{(\nabla ^nx)(\nabla ^n\dot{x})}
+\frac{1}{a^2}\overline{(\nabla ^nx)(\nabla ^{n+2}x)}\\
&&-(\alpha\varphi _{\text{h}}^2+m_\chi ^2)
\overline{(\nabla ^nx)(\nabla ^nx)}\Big]\\
\nonumber &=& 2\overline{y_n}-3H\dot{\overline{x_n}}
-2(\alpha\varphi _{\text{h}}^2+m_\chi ^2)\overline{x_n}\\
&&+(2/a^2)\left[\overline{\mbox{div}((\nabla ^{n+1}x)
(\nabla ^nx))}-\overline{(\nabla ^{n+1}x)^2}\right]\,.
\end{eqnarray*}
The average of the divergence is a surface integral divided by $V$ and
vanishes for $V{\to}\infty$ whence \vskip -0.5\baselineskip
\begin{equation}                                             \label{11}
\ddot{\overline{x_n}}=2\overline{y_n}
-2\overline{x_{n+1}}/a^{2}-3H\dot{\overline{x_n}}
-2(\alpha\varphi _{\text{h}}^2+m_\chi ^2)
\overline{x_n}\,.
\end{equation}
Analogously one obtains
\begin{equation}                                             \label{12}
\dot{\overline{y_n}}=-6H\overline{y_n}-\dot{\overline{x_{n+1}}}/a^{2}
-(\alpha\varphi _{\text{h}}^2+m_\chi ^2)\dot{\overline{x_n}}\,.
\end{equation}
For $n{=}0,1,2,\dots$ Eqs.~(\ref{11})--(\ref{12}) together with
(\ref{5}) and (\ref{fried}) provide an infinite set of equations
coupled together in each order $n$ through a coupling term of order
$n{+}1$.  The coupling terms are multiplied by the factor $1/a^{2}$
that, in an expanding universe, is getting smaller and smaller with
time. In addition, it appears plausible that they will be continually
diminished by smoothing effects of friction. Therefore, it can be
expected that truncation by neglecting the coupling terms to a low
order $n$ will yield good approximations. This can be tested
numerically by comparing results obtained from truncation to different
orders (see Appendix).  It turned out that keeping the $n{=}0$ terms
only is already good enough. With $x_0{=}x^{2}$ and
$y_0{=}\dot{x}^{2}$, the equations obtained this way are \vskip
-\baselineskip
\begin{eqnarray}
  \ddot{\varphi}_{\text{h}}+3H\dot{\varphi}_{\text{h}}
  +\alpha\left(\overline{x_0}-\chi _c^2\right)\varphi_{\text{h}}
  +\lambda\varphi _{\text{h}} ^3&=&0\,,\label{13}\\
  \ddot{\overline{x_0}}+3H\dot{\overline{x_0}}
  +2(\alpha\varphi _{\text{h}}^2+m_\chi ^2)\overline{x_0}
  -2\overline{y_0}&=&0\,,\label{14}\\
  \dot{\overline{y_0}}+6H\overline{y_0}
  +(\alpha\varphi _{\text{h}}^2+m_\chi
  ^2)\dot{\overline{x_0}}&=&0\,,\label{15}
\end{eqnarray}
\vskip -1.4\baselineskip
\begin{eqnarray}
  H^{2}+\frac{k}{a^2} &=&\frac{8\pi(\overline{\rho_{\chi 0}}
    +\rho_{\varphi 0})} {3m_{\text{P}}^2}\,,\label{16}\\
  \rho_{\varphi 0}&=&\frac{\dot{\varphi _{\text{h}}}^2}{2}
  +V_0-\alpha\chi_c^2\frac{\varphi_{\text{h}}^2}{2}+\lambda
  \frac{\varphi_{\text{h}}^4}{4}\,,\label{17}\\
  \overline{\rho_{\chi 0}}&=&\frac{\overline{y_0}}{2}
  +\alpha\frac{\varphi _{\text{h}}^2\overline{x_0}}{2}
  +m_{\chi}^{2}\frac{\overline{x_{0}}}{2}\,,\label{18}\\
  \dot{a}&=&Ha\,.\label{19}
\end{eqnarray}
Note that due to the presence of
$\overline{x_0}=\overline{\tilde{\chi}^{2}}$ and
$\overline{y_0}=\overline{\dot{\tilde{\chi}}^{2}}$ the system still
contains effects of the inhomogeneity of $\chi(\vec{x},t)$, in spite
of the neglection of all gradient terms in the order $n=0$.

$\varphi_{\text{h}}$ carries out damped oscillations in the
neighborhood of a minimum of the potential $V$ and plays the role of a
driver. Eq.~(\ref{14}) for $\overline{x_0}$ has properties similar to
the Mathieu equation, the term $2\alpha\varphi _{\text{h}}^2
\overline{x_0}$ leading to parametric resonance. As a result of this
and the nonlinearities of the system, chaotic solutions must be
expected.

In the early phase of the universe when the total density
$\rho_{0}{=}\rho_{\varphi 0}{+}\overline{\rho_{\chi 0}}$ is very
large, the term $k/a^{2}$ in Eq.~(\ref{16}) can be neglected.
Calculations for low densities will be restricted to the case $k{=}0$.
Thus for the purposes of this paper Eqs.~(\ref{13})--(\ref{18}) build
a closed subset independent of the quantity $a$. With $k{=}0$,
\begin{eqnarray}
  \beta&=&\sqrt{\frac{8\pi}{3}}\,,\quad
  X=\frac{\overline{x_{0}}}{m_{\text{P}}^{2}}\,,\quad
  \Phi=\frac{\varphi_{\text{h}}}{m_{\text{P}}}\,,\quad
  \tau=m_{\text{P}}t\,,\qquad                       \label{neuvara}\\
  \frac{\rho_{\varphi 0}}{m_{\text{P}}^{4}}&\to& \rho_{\varphi}\,,
  \quad
  \frac{\overline{\rho_{\chi 0}}}{m_{\text{P}}^{4}}\to \rho_{\chi}\,,\quad
  \frac{V_{0}}{m_{\text{P}}^{4}}\to V_{0}\,,\quad
  \frac{\chi_{\text{c}}}{m_{\text{P}}}\to \chi_{\text{c}}\label{neuvarb}
\end{eqnarray}
and the definitions $U{=}\dot{\Phi}(\tau)$, $Z{=}\dot{X}(\tau)$,
Eqs.~(\ref{13})--(\ref{18}) for the variables $\varphi_{\text{h}}$,
$\overline{x_{0}}$ and $\overline{y_{0}}$ are transformed into the set
of autonomous first order differential equations
\begin{eqnarray}
  \dot{\Phi}(\tau)&=&U\,,\label{19*}\\
  \dot{U}(\tau)&=&-\left[3HU+\alpha (X-\chi_{\text{c}}^{2})
  \Phi +\lambda \Phi^{3}\right]\label{20}\\
  \dot{\rho}_{\chi}(\tau)&=&-6H\rho_{\chi}+3 \alpha H X
  \Phi ^{2}+\alpha X \Phi U\label{21}\\
  \dot{X}(\tau)&=&Z\label{22}\\
  \dot{Z}(\tau)&=&-3HZ+4\rho_{\chi}-4\alpha X \Phi^{2}\label{23}
\end{eqnarray}
with
\begin{equation}                                            \label{24}
  H{=}\beta\sqrt{\rho_{\varphi}{+}\rho_{\chi}}\,,\quad
  \rho_{\varphi}{=}V_{0}{+}\frac{U^{2}}{2}{-}\frac{\alpha}{2}
  \chi_{\text{c}}^{2}\Phi^{2}{+}\frac{\lambda}{4}\Phi^{4}
\end{equation}
for the dimensionless variables $\Phi$, $U$,  $\rho_\chi$, $X$
and $Z$.  Once the system is solved, $a(\tau)$ is obtained from
\begin{equation}                                             \label{25}
  a=a_{0}\exp{\Big(\beta\int_{\tau_{0}}^{\tau}\!\!
    \sqrt{\rho_{\varphi}(\tau')+\rho_{\chi}(\tau')}\;d\tau'\Big)}\,.
\end{equation}

\section{Conditions on the parameters of the problem}

1.~As is usual in modern cosmology, we shall assume that all
quantities are restricted by the corresponding Planck quantities. For
the (normalized) fields this means $\varphi {{}_{\textstyle <} \atop
  \textstyle \sim}1$ and $\chi{{}_{\textstyle <} \atop \textstyle
  \sim}1$, and for the (normalized) potential $V{{}_{\textstyle <}
  \atop \textstyle \sim}1$.

2.~Presently the energy densities of the matter and inflaton field are
almost equal and very small compared to $1$.  On the assumption that
the corresponding fields are close to a value where the potential has
a minimum, the latter must be very small. In case (\ref{4*}) the
minimum is zero.  In order to avoid fine tuning we assume that this is
also true in case (\ref{4}). In this case, the minimum is obtained for
$\chi{=}0$, $\varphi^{2}{=}\alpha \chi _c^2/\lambda$ and has the value
$V_{\text{min}}{=}V_{0}{-}\alpha \chi _c^2/(4\lambda)$ whence $\alpha
=4\lambda V_{0}/\chi _c^2$ from $V_{\text{min}}{=}0$.

3. A further condition on the parameters of the problem originates
from the requirement that the observed inhomogeneities of the universe
evolve from quantum fluctuations through inflation.  From this, for
the potential $V{=}\lambda \varphi^{4}/4$ in \cite{linde} the
condition $\lambda{\approx} 10^{-14}$ was derived.  Since matter is
getting extremely diluted through inflation, during this process
$\chi{\approx} 0$ whence in case (\ref{4*}) $|\chi|{\ll}|\varphi|$,
$V{\approx} \lambda \varphi^{4}/4$ and thus $\lambda{\approx}
10^{-14}$.  During inflation $\chi{\approx} 0$ holds also for
Eq.~(\ref{4}) whence $V {\approx} V_0{-}\alpha\chi
_c^2\varphi^2/2{+}\lambda\varphi ^4/4$. In this case, a calculation
following the lines of \cite{linde} that, for reason of brevity,
cannot be presented here yields $\lambda {\approx}2{\cdot} 10^{-17}$
for $V_{0}{=}1.25{\cdot}10^{-18}$.

\begin{figure}[tb]
  \includegraphics[width=\columnwidth]{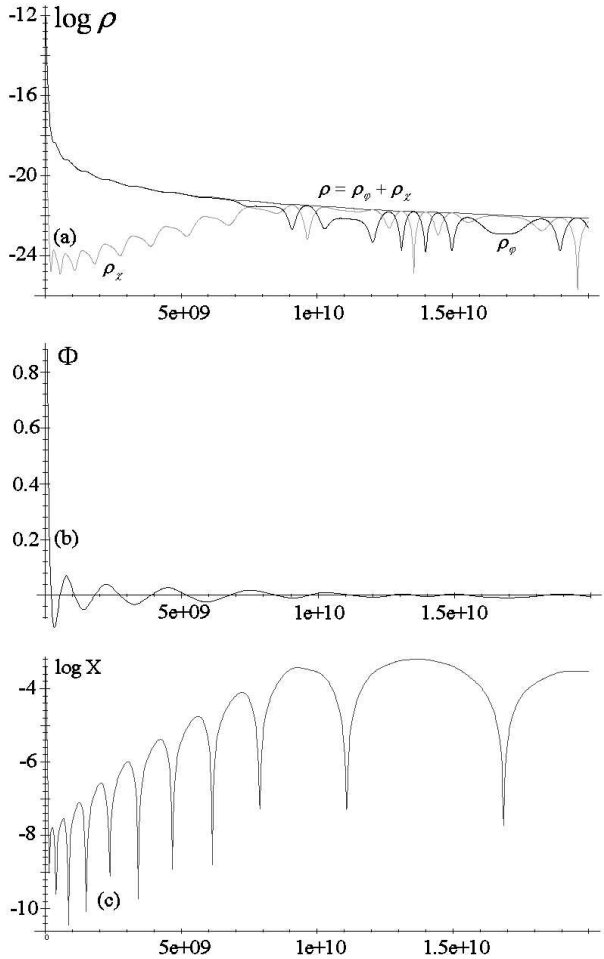}
   \caption{\label{pic:4} (a) Time evolution of $\rho_{\varphi}$,
     $\rho_{\chi}$ and $\rho$ during slow roll and first reheating,
     starting with $\rho
     _{\chi}{=}10^{{-}15}{\ll}\rho _{\varphi}{=}0.15 {\cdot}
     10^{{-}11}$. Below: Evolution of the inflaton field
     $\Phi$ (b) and matter field $X$ (c) over
     the same time interval.}
\end{figure}

\begin{figure}[tb]
  \includegraphics[width=\columnwidth]{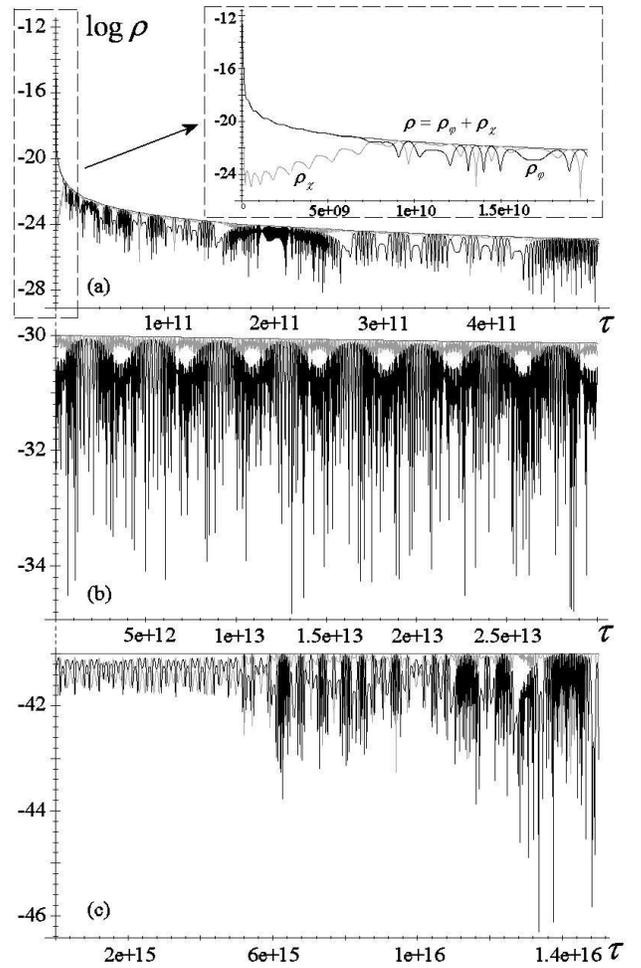}
   \caption{\label{pic:1} Time evolution of $\rho_{\varphi}$,
     $\rho_{\chi}$ and $\rho$ in different time intervals with
     different time scales, $\tau$ being reset to $0$ at the beginning
     of each: (a) before inflation, start with $\rho
     _{\chi}{=}10^{{-}15}{\ll}\rho _{\varphi}{=}0.15 {\cdot}
     10^{{-}11}$ (the same as in Fig.~\ref{pic:4}), (b) much later,
     start with $\rho _{\chi}{=}10^{{-}30}{\gg}\rho
     _{\varphi}{=}10^{{-}34}$, (c) start with $\rho _{\chi}{=}\rho
     _{\varphi}{=}5{\cdot}10^{-41}$.}
\end{figure}

\section{Inflaton field model}

The case of the parabola shaped potential (\ref{4*}) is essentially
the scenario of Linde's chaotic inflation \cite{linde}. It is known
that for the small value $\lambda\approx 10^{-14}$ an extreme
inflationary expansion is obtained if the total density $\rho$ starts
at about the Planck density \cite{linde}.  Therefore for numerical
calculations a much lower initial value $\rho(0){\approx} 0.16{\cdot}
10^{{-}12}$ was chosen, leading to an inflationary enhancement of the
scale factor $a$ by 35 orders of magnitude only. In order to obtain
inflation, $p_{\varphi 0}= \dot{\varphi}^{2}_{\text{h}}/2-
V_{\varphi}(\varphi_{\text{h}})<0$ must be satisfied which was
achieved by setting $\dot{\varphi}_{\text{h}}(0)=0$ or $U(0)=0$. With
this, according to Eqs. (\ref{24})
\begin{displaymath}
  \Phi(0)=\left(\frac{4\rho_{\varphi}(0)}{\lambda}\right)^{1/4}\,.
\end{displaymath}

For matter, between pressure and density a relation $\overline{p_{\chi
    0}}=\gamma \overline{\rho_{\chi 0}}$ holds where $\gamma=1/3$ in
the highly relativistic regime and $\gamma\ll 1/3$ in the non
relativistic regime. With Eq.  (\ref{17}) for $m_{\chi}=0$ and
$\overline{p_{\chi 0}}=\overline{y_0}/2 -\alpha\varphi
_{\text{h}}^2\overline{x_0}/2$ this relation leads to
$\overline{y_0}=(1+\gamma)\alpha\varphi _{\text{h}}^2\overline{x_0}
/(1-\gamma)$ and $\overline{x_0}=(1-\gamma)\overline{\rho_{\chi 0}}/
(\alpha\varphi _{\text{h}}^2)$ or, expressed in the variables
introduced in Eqs. (\ref{neuvara})--(\ref{neuvarb}),
\begin{eqnarray}                                  \label{anfx}
  X&=&\frac{(1{-}\gamma)\rho_{\chi}}{\alpha\,\Phi^{2}}\,,\\
  Z&=&\frac{1+\gamma}{1-\gamma}\alpha\Phi X\,.
\end{eqnarray}
These conditions were used as initial conditions for $X$ and $Z$. The
exact value given initially to $\gamma$ turned out to be of no
importance in the long run, so we used $\gamma=1/3$.

For $\rho_{\chi}$ various initial conditions were considered, from
essentially no matter to equipartition between matter and inflaton
field energy. The results are discussed in the following.

$\rho _{\chi}$ is first dramatically reduced by inflation
(Fig.~\ref{pic:4} and Fig.~\ref{pic:1} (a), insert). When $\Phi$
starts to oscillate about $\Phi{=}0$, reheating sets in and
$\rho_{\chi}$ is raised. The frequency of the heating oscillations is
coupled to the one of the oscillations of $\Phi(t)$ and is
approximately twice as large. The largest $\rho_{\chi}$ reached after
reheating is a function of $\alpha$.  The condition that
$\rho_{\chi}=\rho_\varphi$ at some point in the evolution of the
Universe yields the constraint $1.2 \,\lambda\, {{}_{\textstyle <}
  \atop \textstyle \sim} \alpha {{}_{\textstyle <} \atop \textstyle
  \sim} 3\,\lambda\,$. For the numerical calculations $\alpha{=}2
\,\lambda$ was used. After $\rho_{\chi}$ and $\rho _{\varphi}$
coincide they start to oscillate in a chaotic manner around each
other.

\begin{figure}[tb]
  \includegraphics[width=\columnwidth]{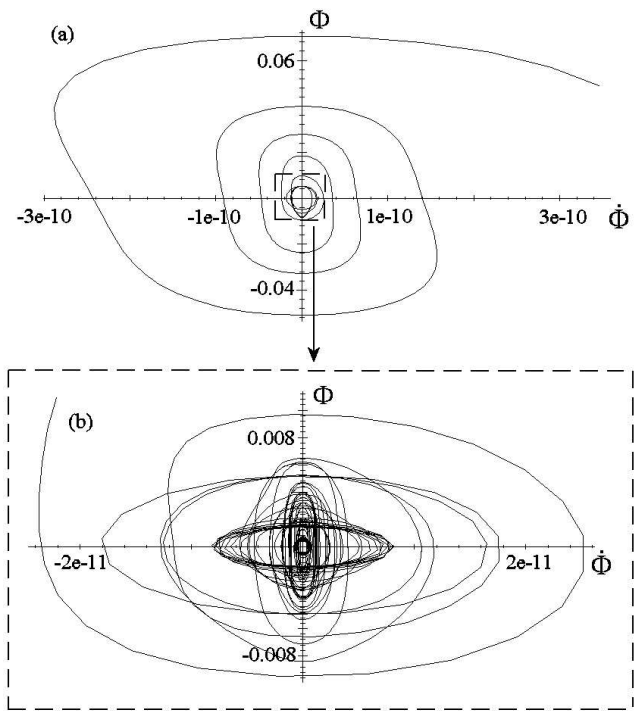}
   \caption{\label{pic:5} Phase portraits of the inflaton-field $\Phi$:
   (a) the same time interval and same initial settings
   as in Fig.~\ref{pic:4} were used,
   only the slow roll phase is omitted; (b) continues where (a) stopped
   and covers the rest of  the time interval used in Fig.~\ref{pic:1} (a) --
   the chaotic nature of the oscillations is easy to see.}
\end{figure}

A phase portrait of $\varphi$ shown in Fig.~\ref{pic:5} exhibits
typical chaotic structure.  Chaotic behavior arises naturally when two
or more scalar fields interact with each other in an expanding
Universe \cite{felder,joras,cornish}, but it is usually of a transient
type. The observed chaotic oscillations in the present model, on the
contrary, are persistent.  Even after quite large excursions
$\rho_{\chi}(t)$ and $\rho _{\varphi}(t)$ always find back to each
other.  This kind of intermittent coincidence holds true for the whole
decay of $\rho{=}\rho _{\chi}{+}\rho _{\varphi}$ from the very large
values before inflation down to very small values like those of today.

\begin{figure}[tb]
  \includegraphics[width=\columnwidth]{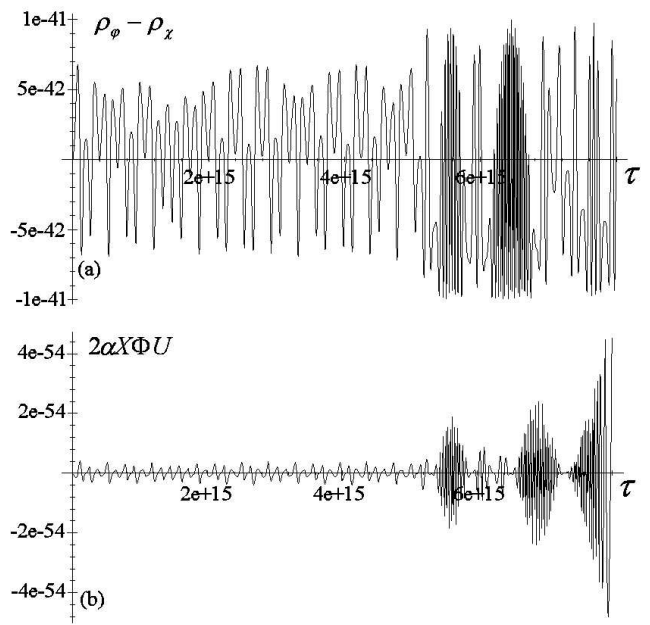}
   \caption{\label{pic:8} Density difference $\rho_{\varphi}(\tau)-\rho_{\chi}(\tau)$
     and driving term $2\alpha X(\tau)\Phi(\tau) U(\tau)$ of equation
     (\ref{driver}). The same settings and initial conditions as in
     figure \ref{pic:1}\,(c) were used, but only about half of the
     time-interval was covered.}
\end{figure}

It can qualitatively be understood as follows. From equations
(\ref{19*})--(\ref{21}) and (\ref{24}) one easily derives
\begin{equation}                                   \label{drph}
 \dot{\rho}_{\varphi}(\tau)=-6H\rho_{\varphi}+(3/2)\lambda
H\Phi^{4}-\alpha X \Phi U
\end{equation}
and
\begin{eqnarray}
\nonumber \frac{d(\rho_{\varphi}-\rho_{\chi})}{d\tau}
&=&-6H(\rho_{\varphi}-\rho_{\chi})
-3\alpha X\Phi^{2}\\
&&+\frac{3}{2}\lambda H\Phi^{4}-2\alpha X\Phi U\,.\label{driver}
\end{eqnarray}
The first term on the right hand side of Eq. (\ref{driver}) is always
leading to a decrease of $\rho_{\varphi}-\rho_{\chi}$. The second term
is $\leq 0$ and the third is $\geq 0$ so the two are counteracting,
and numerically their sum turns out to be smaller than the first term.
The last term is oscillatory and is dominant according to the
numerical evaluation. It is a forcing term that leads to oscillations
of $\rho_{\varphi}-\rho_{\chi}$ because according to Eqs. (\ref{21})
and (\ref{drph}) one half of it increases $\rho_{\varphi}$ when the
other half decreases $\rho_{\chi}$ and vice versa. A demonstration of
this behavior is given in Figs.~\ref{pic:8} which show results for
$\rho_{\varphi}(\tau)-\rho_{\chi}(\tau)$ and $2\alpha
X(\tau)\Phi(\tau) U(\tau)$.

\begin{figure}[tb]
  \includegraphics[width=\columnwidth]{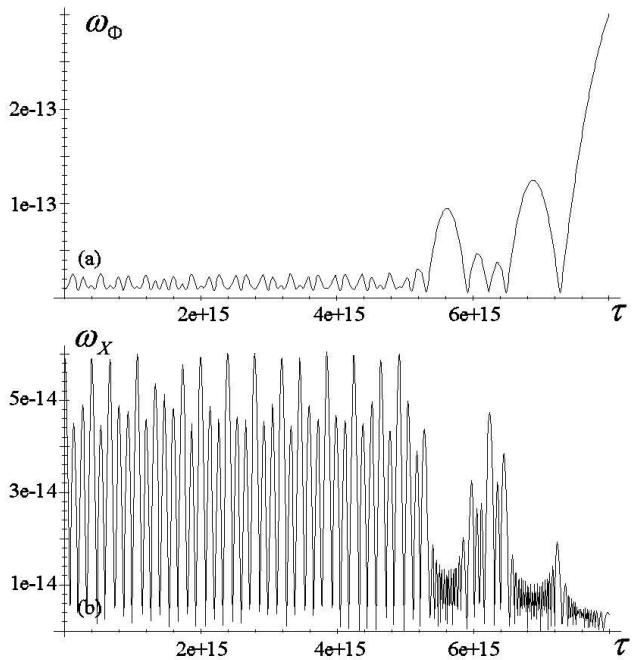}
   \caption{\label{pic:7} Approximate frequencies $\omega_\Phi\approx\sqrt{\alpha X}$ and
     $\omega_X\approx 2\left|\Phi\right|\sqrt{\alpha }$, for the same
     settings and initial conditions as used in figure
     \ref{pic:1}\,(c), but only about half of the time-interval was
     covered.}
\end{figure}

Some insight into the oscillations involved in the process is obtained
by the following consideration.  For $\chi_{\text{c}}=0$ combining
Eqs. (\ref{19*}) and (\ref{20}) yields
\begin{equation}
  \ddot{\Phi}(\tau)=-3H\dot{\Phi}-\alpha X\Phi-\lambda \Phi^{3}\,.
\end{equation}
This is an equation for the oscillations of $\Phi$ in which $-(\alpha
X\Phi+\lambda \Phi^{3})$ is the driving force and $-3H\dot{\Phi}$ is a
friction term. After a few oscillations through $\Phi=0$ the dominant
term becomes $-\alpha X \Phi$, and from
\begin{displaymath}
  \ddot{\Phi}(\tau)\approx-\alpha X\Phi
\end{displaymath}
the oscillation frequency
\begin{equation}                                       \label{omphi}
  \omega_{\Phi}\approx\sqrt{\alpha X}
\end{equation}
is obtained.  From Eqs. (\ref{22})--(\ref{23}) we get
\begin{displaymath}
  \ddot{X}(\tau)=-3H\dot{X}+4\rho_{\chi}-4\alpha X\Phi^{2}\,.
\end{displaymath}
Again this is an equation for oscillations, this time of $X(\tau)$,
with $-4\alpha X\phi^{2}$ being the driving term, and an approximate
expression for the frequency of the oscillations is
\begin{equation}                                       \label{omX}
  \omega_{X}\approx 2\left|\Phi\right|\sqrt{\alpha }\,.
\end{equation}
According to the numerical calculations, on the average
\begin{displaymath}
  \frac{\omega_{X}}{\omega_{\Phi}}\approx \frac{2\left|\Phi\right|}{\sqrt{X}}
\end{displaymath}
is smaller than 1 whence the oscillations of $X$ are slower than those
of $\Phi$. In consequence, in the short run the frequency
$\omega_{\Phi}\approx\sqrt{\alpha \chi}$ of the $\Phi$ oscillations
varies between smaller and larger values as can be clearly seen in
Figure~\ref{pic:7} and is also reflected in Fig.~\ref{pic:8}. In the
long run $X$ decreases like all other variables of the system of Eqs.
(\ref{19*})--(\ref{24}). Therefore, as time goes on, the average
frequency of the $\Phi$ oscillations and through this also that of the
oscillations of $\rho_{\varphi}$ and $\rho_{\chi}$ around each other
is slowly decreasing.

Obviously the chaotic nature of the system is still quite structured,
because roughly the two frequencies $\omega_{X}$ and $\omega_{\Phi}$
remain visible as organizing factors. It is well known that in linear
systems with two incommensurable frequencies ergodic phase portraits
are obtained. It appears that the chaoticity of the present system is
caused by a permanent change of the ratio $\omega_{\Phi}/\omega_{X}$.
In this process, due to the overwhelming abundance of irrational
numbers, the change between different measures of incommensurability
leads to a certain amount of unpredictability.

Because it was not possible to calculate the full evolution from
inflation to the present state, the time-averaged coincidence of
$\rho_{\varphi}$ and $\rho_{\chi}$ claimed above was tested piecewise
in more than ten different $\rho$-regimes from very large to very
small, starting with many different values of the ratio $\rho
_{\chi}/\rho _{\varphi}$ in each.  Fig.~\ref{pic:1} shows 3 typical
examples.

For the evolution of $\varrho$ after inflation the approximate
validity of $\varrho a^{\nu}{=}\text{const}$ was found with
$\nu{\approx}4$ in (a), $\nu{\approx}3.9$ in (b) and $\nu{\approx}4$
in (c) etc., i.e. the usual relativistic equation of state, $\varrho
a^{4}{=}\text{const}$, is a good approximation to the evolution of the
(rather smooth) total density $\varrho$.

\section{Remarks on the Coincidence Problem}

In Ref. \cite{dodelson} an idea for a solution of the coincidence
problem \cite{caldwell} was elaborated: The inflaton field energy,
called dark energy there, has periodically dominated in the past,
giving a finite probability to its observed preponderance today. The
oscillations are achieved by sinusoidal modulations of an
exponentially decaying potential for the scalar field representing
dark energy, i.e. by the assumption of a periodic forcing.

Our model leads to a similar behavior, only that the oscillations are
not purely periodic but slightly chaotic.  In contrast to Ref.
\cite{dodelson} they are not implied by an imposed forcing but caused
by the oscillations of the inflaton field $\Phi$ around $\Phi=0$ and
its interaction with the matter field $\chi$.  The mechanism is the
following: The oscillations of $\Phi(t)$ cause heating oscillations by
which again and again the energy density of matter is temporarily
raised above the energy density of the inflaton field. Since with
decreasing amplitude of the $\Phi$ oscillations also the frequency is
getting smaller and smaller, the periods with a preponderance of
either matter or inflaton field become increasingly longer.

In our model the alternating preponderance of matter and inflaton
field was observed over a density range from $10^{-12}$ to $10^{-115}$
with the same set of parameters $\alpha$ and $\lambda$. However, in
spite of its continuous increase at low densities the average period
of preponderance is still too small by orders of magnitude. With
$\lambda=10^{-120}$ instead of $\lambda=10^{-14}$ a reasonable period
(small fraction of the life time of the universe) can be achieved, but
with this small value the behavior at large densities is no longer the
needed reheating scenario, which was found only down to values
$\lambda{{}_{\textstyle >} \atop \textstyle \sim} 10^{-40}$. At first
glance it would appear that a potential $V(\Phi)$ that decays like
$10^{-14}\Phi ^{4}$ at large $\Phi$ and like $10^{-120}\Phi ^{4}$ at
small $\Phi$ would lead to a solution. It turns out, however, that an
effective reheating at large densities requires a large value of
$\lambda$ even around $\Phi=0$.

At second glance a dynamic $\lambda=\lambda(t)$ would appear to do it,
and $\lambda(t)\sim H^{2}(t)$ would look like a good choice, coupling
the potential directly to the energy of the gravitational field.
However, in a reasonable dynamical evolution pressure and density
should be connected by $\overline{p_{\chi 0}}=\gamma
\overline{\rho_{\chi 0}}$ and in consequence, Eq. (\ref{anfx}) should
hold. From this and \eqref{omphi} we get
\begin{displaymath}
  \omega_{\Phi}\approx\frac{\sqrt{\rho_\chi}}{|\Phi|}\approx
  \frac{H}{|\Phi|}
  \geq\frac{H}{\mbox{max}\,|\Phi|}.
\end{displaymath}
While at later times $\omega_{\Phi}\approx H$ would have the desired
order of magnitude because $1/H$ is approximately the age of the
universe, the small factor $1/\left |\Phi\right |$ spoils any attempt
to obtain reasonably small frequencies.

From our studies it can be concluded, that the oscillations of the
inflaton field and its interaction with matter can keep the energy
densities of the two close together over the full range of densities
from inflation to present day densities, with alternating
preponderance of one of the two. Reasonable times for the duration of
the preponderance at present day densities cannot be obtained from our
model. So oscillations of the inflaton field and periodic reheating of
matter through a coupling between matter and inflaton field may be one
clue to the solution of the coincidence problem but cannot provide a
full solution of it. The simple model for describing matter adopted in
this paper is certainly not appropriate for later stages in the
evolution of the universe, and a better adapted model might improve
the situation.

\section{Higgs field model}
\label{sechfm}

\begin{figure}[tb]
  \begin{center}
   \includegraphics[width=0.45\columnwidth]{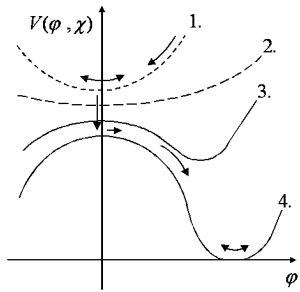}\\
   \caption{\label{pic:2} Potential (\ref{4}) for 4 different
     values of $\overline{x_{0}}\chi$ that decrease in the shape
     sequence 1,2,3,4. During pre-inflation shapes 1 and 2 are active,
     during inflation shape 3. 4 is the final shape when matter is
     already very diluted.}
   \end{center}
 \end{figure}
\begin{figure}[t]
  \begin{center}
   \includegraphics[width=0.9\columnwidth]{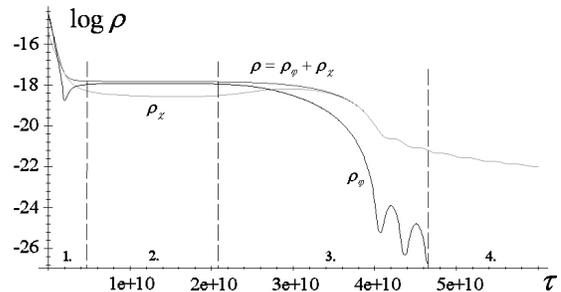}\\
   \caption{\label{pic:3} Time evolution of $\rho_{\varphi}$ and
     $\rho_{\chi}$ for the potential (\ref{4}) that evolves according
     to Fig.~\ref{pic:1}.}
 \end{center}
\end{figure}
For the potential (\ref{4}) calculations were done with
$V_0{=}1.25{\cdot} 10^{-18}$, $\lambda{=}2{\cdot} 10^{{-}17}$,
$\alpha{=}\lambda /10$, $\chi _c{=}\sqrt{5}$, $m_{\chi}^2{=}1{\cdot}
10^{{-}19}$ and with the initial values $\rho _{\chi}{=}7.5{\cdot}
10^{{-}16}$, $\rho _{\varphi}{=}3 {\cdot} 10^{{-}15}$ of the
densities.  First the potential assumes shape 1 of Fig.~\ref{pic:2},
and $\varphi$ starts at a value for which $V$ is well above its
minimum at $V{=}0$.  As $\varphi$ rolls slowly towards this,
$\rho_{\chi}$ decreases whence $V$ gradually assumes a flatter shape
like 2. When $\varphi$ reaches the region around $\varphi{=}0$,
$\rho_{\chi}$ may be enhanced through oscillations of $\varphi$ in the
potential well but is then again diluted by ongoing inflation. Finally
$\overline{x_{0}}{=}\overline{\tilde{\chi}^{2}}$ becomes smaller than
$\chi_{c}^{2}$ and $V(\varphi)$ assumes a sombrero-like shape as 3.
Then $\varphi$ starts to roll downhill towards a new minimum appearing
at a $\varphi{\neq} 0$. Oscillations around this lead to an increase
of $\rho_{\chi}$. Due to the change of shape of the potential the
oscillation frequency changes and thus parametric resonance occurs at
many different frequencies, leading to an especially effective
reheating. Asymptotically, shape 4 will be assumed and $\varphi$ will
settle at the minimum $V{=}0$ obtained from Eq.~(\ref{4}) for
$\chi{=}0$.  The dynamics can be separated into four qualitatively
different steps (Fig.~\ref{pic:3}).  1. Pre-inflation: Slow roll in a
potential of shape 1-2 towards $\varphi{=}0$ with possible reheating
through oscillations around $\varphi{=}0$. 2. Inflation: Slow roll
from $\varphi{=}0$ in a potential of shape 3 towards a new minimum at
$\varphi{\neq} 0$. 3.  Reheating: Oscillations around the new minimum.
4. Friedmann-like evolution after the settlement of $\varphi$ at the
new minimum.  These four steps are found for a wide variety of
parameters and initial conditions.

A model combining the properties of the two cases considered, chaotic
oscillations with alternating dominance of matter and inflaton field
as well as dynamic symmetry breaking, should be obtainable by
employing two inflaton fields of type $\varphi=\varphi _{\text{h}}(t)
+\varepsilon \tilde{\varphi}(\vec{x},t)$, one moving in a parabola-
and the other in a sombrero-like potential, and both being coupled to
the matter field.

\section{Conclusions}

The main result of this paper consists in a set of equations for the
interaction between a scalar inflaton field, a scalar matter field and
the gravitational field, obtained by averaging processes. Due to the
averaging no specific assumptions about the inhomogeneities of the
matter field like specifying any initial conditions must be made. The
set of equations can be applied to pre-inflation, inflation and the
reheating after inflation. It contains the heating effects of
parametric resonance and is nevertheless rather easy to handle.
However, it should be noted that the presented approach is neither
covariant, nor gauge invariant. Therefore one has to take great care
to compute only physically meaningful quantities, as in the studied
cases.

Processes like the dynamic symmetry breaking of a Higgs field, the
theoretical description of which usually requires quite some effort,
are easily obtained and understood. An attempt to gain some insight
into the coincidence problem was made.  Repeated periods of matter
reheating can keep the energy densities of the inflaton and the matter
field close together for all times until today and lead to an
alternating dominance of one of them. The preponderance of the
inflaton field observed today would thus be a question of chance and
be preceded by a preponderance of matter. However, the specific model
employed for describing this process did not yield reasonable
durations of preponderance over the full evolution from inflation
until today.  We therefore conclude that it may provide one clue to
the understanding of the problem while other clues may still be
missing.

\vspace \baselineskip

{\bf Acknowledgment.} This work was partly support\-ed by Deutsche
Forschungsgemeinschaft.  E.~R. expresses his gratitude for hospitality
and support of the Physics Dept. at Brown University, Providence,
R.~I., USA, received in the initial phase of this work.  Valuable
discuss\-ions with R.  Brandenberger are gratefully acknowledged.
\section{Appendix}

\begin{figure}[t]
  \includegraphics[width=\columnwidth]{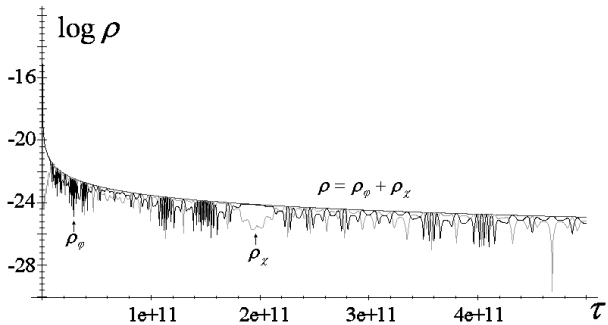}
   \caption{\label{pic:6} Time evolution of $\rho_{\varphi}$,
     $\rho_{\chi}$ and $\rho$ computed to order $n=1$. The same time
     interval and initial conditions as in Fig.~\ref{pic:1} (a) were
     used, except for $a$ and $\overline{x_1}$ which was set to
     $a_{i}\equiv m_{\text{P}}^{-1}$ and $\overline{y_0}a_{i}^2/10$
     respectively.  Comparison with Fig.\ref{pic:1} (a) shows a
     similar evolution of the densities.}
\end{figure}
The quality of the approximation provided by equations
(\ref{19*})--(\ref{24}) can easily be tested by truncation to higher
orders.  Truncating to order $n=1$ enhances our set
(\ref{19*})--(\ref{24}) by two equations and the two variables $x_{1}$
and $y_{1}$ as
\begin{eqnarray*}
\ddot{\overline{x_0}}&=&2\overline{y_0}
-2\overline{x_{1}}/a^{2}-3H\dot{\overline{x_0}}
-2(\alpha\varphi _{\text{h}}^2+m_\chi ^2)
\overline{x_0}\,,\\
\dot{\overline{y_0}}&=&-6H\overline{y_0}-\dot{\overline{x_{1}}}/a^{2}
-(\alpha\varphi _{\text{h}}^2+m_\chi ^2)\dot{\overline{x_0}}\,;\\
\ddot{\overline{x_1}}&=&2\overline{y_1}
-3H\dot{\overline{x_1}}
-2(\alpha\varphi _{\text{h}}^2+m_\chi ^2)
\overline{x_1}\,,\\
\dot{\overline{y_1}}&=&-6H\overline{y_1}
-(\alpha\varphi _{\text{h}}^2+m_\chi ^2)\dot{\overline{x_1}}\,.
\end{eqnarray*}
In addition, one needs to compute the time evolution of the scale
factor $a$ by means of equation (\ref{19}) simultaneously, since it
enters the equations now.

In view of their chaotic character a detailed comparison of the fine
structure of the solutions obtained with truncation to order $n=0$ and
to order $n=1$ is not meaningful. However, the average behavior in the
long run must be comparable, and indeed all computations yielded the
same qualitative and essentially the same quantitative behavior -
compare Fig.~\ref{pic:1} (a) with Fig.~\ref{pic:6} as an example.  No
crucial dependence on the initial values of $\overline{x_1}$ and
$\overline{y_1}$ was observed, so we choose
$\overline{x_0}=(1-\gamma)\rho_\chi/(\alpha\Phi ^2)$,
$\overline{y_0}=(1+\gamma)\rho _\chi/2$,$\overline{x_1}=(1+\gamma)\rho
_\chi/2$ and $\overline{y_1}=0$, with $\gamma=1/3$ in
Fig.~\ref{pic:6}. As an initial value for the scale factor we choose
$a=m_{\text{P}}^{-1}$, corresponding to a Planck sized patch of the
Universe.

By the same token the Higgs field model of section \ref{sechfm} was
checked up to second order and no essential difference of the long run
behavior was found either.

From this comparison it can be concluded that truncation to the order
$n=0$ already provides a very good approximation.


\begin{thebibliography}{10}

\bibitem{linde} A. Linde,
{\it Particle Physics and Inflationary Cosmology}
 (Harwood, Chur, Switzerland, 1990).

\bibitem{lyth} D. H. Lyth and A. Riotto,
 Phys. Rep. {\bf 314}, 1 (1999).

\bibitem{brand3} R. Brandenberger, in
{\it Large Scale Structure Formation} edited by R. Mansouri
and R. Brandenberger
(Kluwer, Dordrecht, 2000).

\bibitem{linde2} L. Kofman, A. Linde,
and A. Starobinsky, Phys. Rev. D {\bf 56}, 3258 (1997).

\bibitem{linde3} P. Greene, L. Kofman, A. Linde,
and A. Starobinsky, Phys. Rev. D {\bf 56}, 6175 (1997).

\bibitem{linde4} L. Kofman, A. Linde,
and A.Starobinsky, Phys. Rev. Lett. {\bf 73}, 3195 (1994).

\bibitem{brand} J. Traschen
and R. Brandenberger, Phys. Rev. D {\bf 42}, 2491 (1990).

\bibitem{brand2} Y. Shtanov, J. Traschen
and R. Brandenberger, Phys. Rev. D {\bf 51}, 5438 (1995).

\bibitem{joras} S.E.Jor\'as, V.H.C\'ardenas, Phys. Rev. D 67, 043501 (2003).

\bibitem{felder}G. Felder and L. Kofman, Phys. Rev. D63 103503 (2001).

\bibitem{liddle} A. L. Liddle and D. H. Lyth, {\it Cosmological
    inflation and Large-Scale Structure}, (Cambridge Univ. Press,
  Cambridge 2000)

\bibitem{raychaurdhuri1} A. K. Raychaudhuri, Phys. Rev. {\bf 98},
    1123 (1955)

\bibitem{raychaurdhuri2} A. K. Raychaudhuri, {\it Theoretical
    Cosmology}, (Clarendon, Oxford, 1979)

\bibitem{balescu} R. Balescu, {\it Transport Processes in Plasmas},
  (Elsevier, Amsterdam, 1988)

\bibitem{cornish}N.J. Cornish and J.J. Levin, Phys. Rev. D
{\bf 58},3022 (1996).

\bibitem{dodelson} S. Dodelson, M. Kaplinghat and E. Stewart,
  Phys. Rev. Lett. {\bf 85}, 5276 (2000)

\bibitem{caldwell} R. Caldwell, R. Dave and P.J. Steinhardt, Phys. Rev. Lett.
    {\bf 80}, 1582 (1998)



\end{thebibliography}
\end{document}